\documentclass[11pt,twoside]{article}

\usepackage{asp2004}
\usepackage{epsf}
\usepackage{graphicx}
\usepackage{lscape}

\begin{document}
\title{Are Wolf-Rayet winds driven by radiation?}    
\author{G\"otz Gr\"afener \& Wolf-Rainer Hamann}   
\affil{Department of Physics, University of Potsdam, D-14469 Potsdam, Germany}    

\begin{abstract} 
  Recent results with the Potsdam Wolf-Rayet (PoWR) models have shown that
  Wolf-Rayet mass loss can be explained by radiative wind driving.  An
  inspection of the galactic WR sample, however, reveals that a significant
  part of the observed WR stars are in conflict with our models. This group is
  chiefly formed by intermediate spectral subtypes. Among the population of
  late-type WN stars we find that the enigmatic WN\,8 subtypes, which are
  well-known for their photometric variations, are in disagreement with our
  models. So are there other driving mechanisms working in these objects?
\end{abstract}


\begin{figure}[p!]
\unitlength 0.99\textwidth
\begin{picture}(0.995,1.252)
\put(0.07,0.852){\parbox[b]{0.91\unitlength}{\includegraphics[width=0.91\unitlength,height=0.39\unitlength, clip, trim=25 25 70 25]{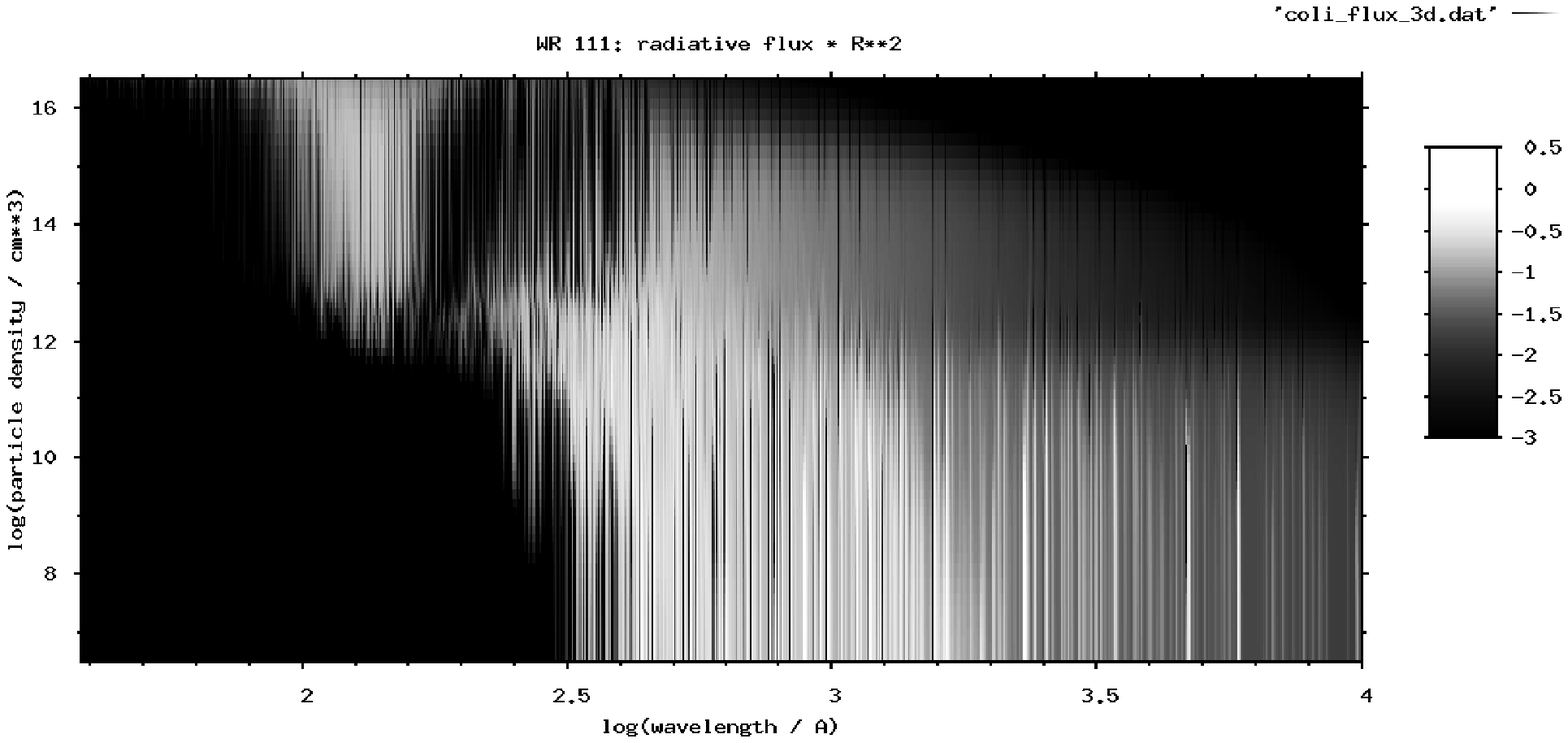}}}
\put(0.07,0.452){\parbox[b]{0.91\unitlength}{\includegraphics[width=0.91\unitlength,height=0.39\unitlength, clip, trim=25 25 70 25]{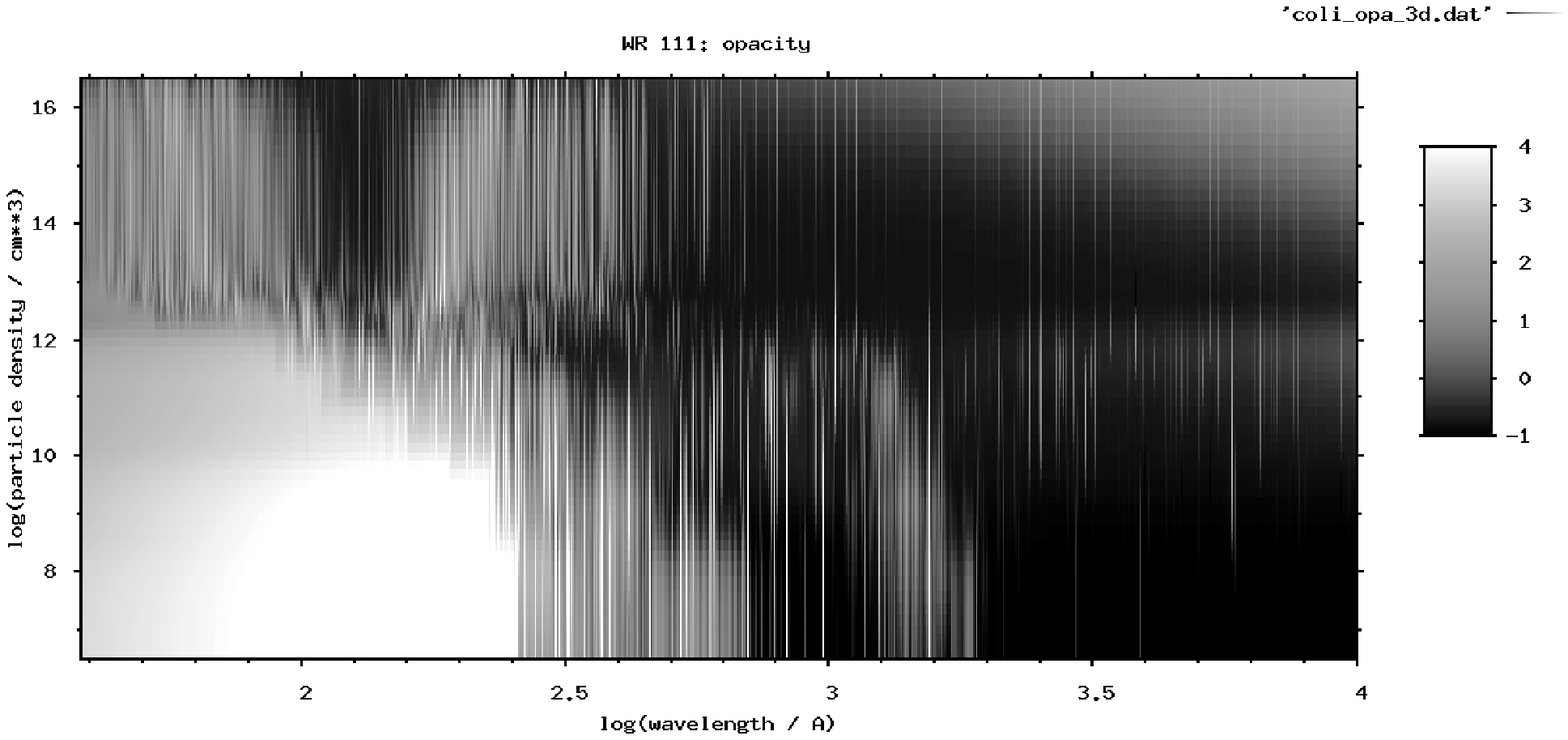}}}
\put(0.07,0.052){\parbox[b]{0.91\unitlength}{\includegraphics[width=0.91\unitlength,height=0.39\unitlength, clip, trim=25 25 70 25]{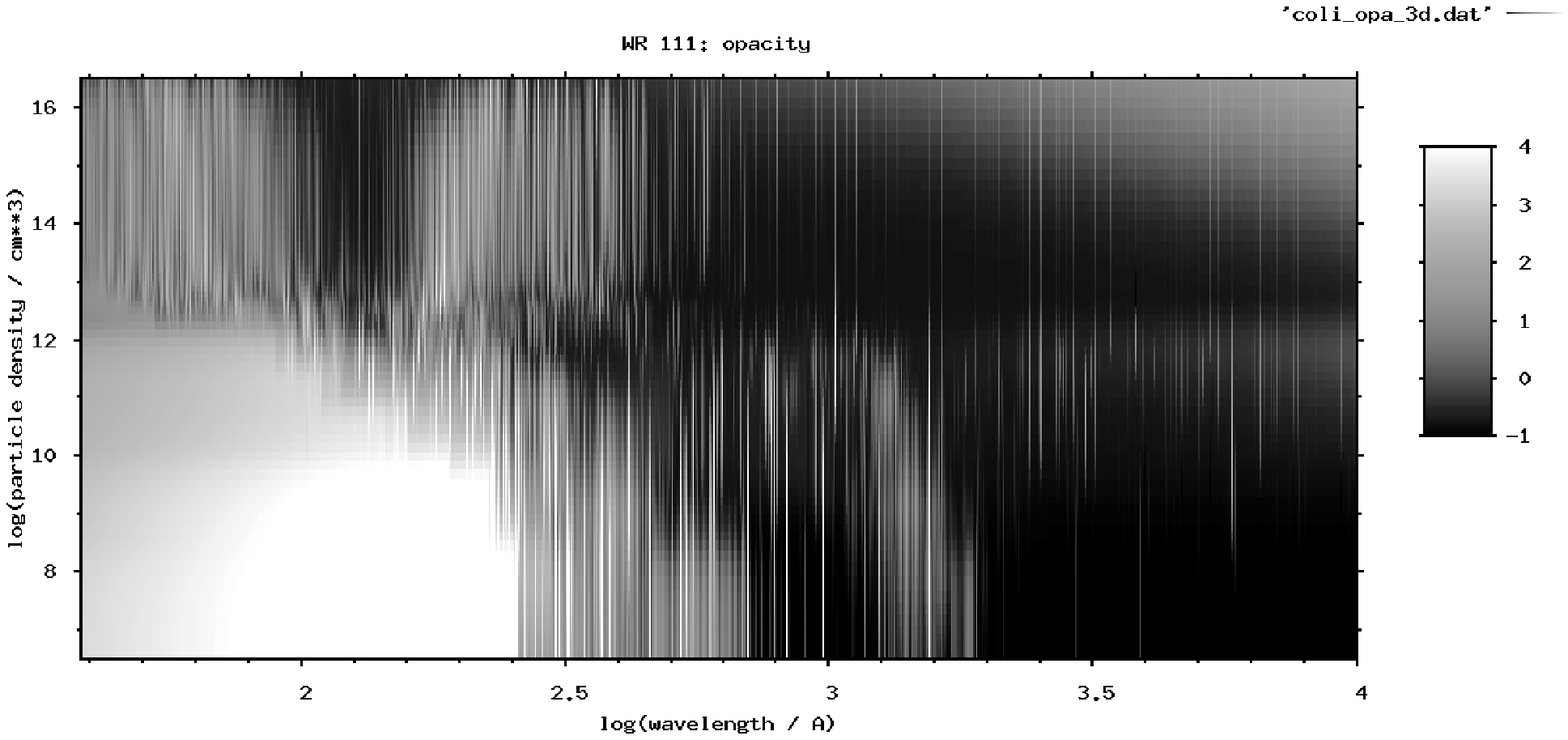}}}
{\footnotesize
\put(0.037,1.213){16}
\put(0.037,1.138){14}
\put(0.037,1.060){12}
\put(0.037,0.982){10}
\put(0.050,0.905){8}
\put(0.00,0.96){\rotatebox{90}{\small $\log(n_{\rm atom}/{\rm cm}^3)$}}
\put(0.037,0.813){16}
\put(0.037,0.738){14}
\put(0.037,0.660){12}
\put(0.037,0.582){10}
\put(0.050,0.505){8}
\put(0.00,0.56){\rotatebox{90}{\small $\log(n_{\rm atom}/{\rm cm}^3)$}}
\put(0.037,0.413){16}
\put(0.037,0.338){14}
\put(0.037,0.260){12}
\put(0.037,0.182){10}
\put(0.050,0.105){8}
\put(0.00,0.16){\rotatebox{90}{\small $\log(n_{\rm atom}/{\rm cm}^3)$}}
\put(0.222,0.03){2}
\put(0.398,0.03){2.5}
\put(0.594,0.03){3}
\put(0.771,0.03){3.5}
\put(0.965,0.03){4}
\put(0.48,0.){\small $\log(\lambda/{\rm \AA})$}
}
\end{picture}
\caption{Self-consistent wind model for an early-type WC star from \citet{gra1:05}:
  radiative flux (top), mass absorption coefficient (middle), and the
  resulting radiative acceleration per frequency interval (bottom) are plotted
  vs.\ wavelength, and wind density (as depth index). The wavelength ranges
  from X-ray to IR and the depth from the inner boundary at $\tau_{\rm
    Ross}=20$ outward to $\approx 1000\,R_\star$.  The opacity distribution
  shows successively recombining continua due to C, O, and He in the FUV, and
  strongly overlapping line opacities in the UV. The radiative acceleration is
  mainly due to Fe-group line opacities absorbing the complex radiative flux.
  \label{fig:3d}
}
\end{figure}

\section{PoWR hydrodynamic model atmospheres}

The Potsdam Wolf-Rayet (PoWR) hydrodynamic model atmospheres combine fully
line-blanketed non-LTE models with the equations of hydrodynamics
\citep[see][]{gra1:05,ham1:03,koe1:02,gra1:02}.  In these models the wind
structure ($\rho(r)$, $v(r)$, and $T(r)$) is computed in line with the full
set of non-LTE populations and the radiation field in the co-moving frame
(CMF). In this numerically rather expensive approach no simplifying
assumptions for the computation of the radiative acceleration $a_{\rm rad}$
are made. In contrast to all previous wind models $a_{\rm rad}$ is
obtained by direct integration instead of making use of the Sobolev
approximation
\begin{equation}
 \label{eq:arad}
  a_{\rm rad} = \frac{1}{c} \int \chi_\nu F_\nu {\rm d}\nu.
\end{equation}
 In this way,
complex processes like strong line overlap or the redistribution of radiation
are automatically taken into account.  The models thus describe the conditions
in WR\,atmospheres in a realistic manner, in particular also at large optical
depth.  Moreover they provide synthetic spectra, i.e.\ they allow for a direct
comparison with observations.

Utilizing these models, we obtained the first fully self-consistent Wolf-Rayet
wind model \citep{gra1:05}, for the case of an early-type WC star.  Moreover
we have recently examined the mass loss from late-type WN stars and its
dependence on metallicity \citep{gra2:06,gra1:06}.  To illustrate the
complexity of the involved processes we show in Fig.\,\ref{fig:3d} the
distribution of the radiative flux, the opacity, and their product (i.e., the
radiative acceleration per frequency interval) within the WC model. It can be
clearly seen that the initial FUV flux, originating from the bottom of the
wind, is blocked by successively recombining continua.  This radiation is
re-emitted at lower frequencies in the form of emission lines, forming the
typical WR emission line spectrum.  The radiative acceleration predominantly
arises from Fe-group line opacities absorbing this radiation. These lines
partly form the observable iron forest short-ward of $1500\,$\AA.  They show a
pronounced ionization structure and contribute by themselves to the complex
'background' flux.  This coexistence of line emission and absorption, in
combination with strong overlap, is the most probable reason for substantial
deviations from the standard theory of line-driven winds \cite[][see also
Bjorkman this volume]{cas1:75}.  In our WR models we tend to detect extremely
small or even negative force multiplier parameters \citep[see][]{gra1:05},
which can be explained by a 'statistical' line locking effect due to the
strong overlap.

In addition to the complex radiative processes we find a strong influence of
wind clumping on the dynamics of O and WR star winds \citep{gra2:03,gra1:05}.
Wind clumping particularly affects the recombination rates because these scale
with $n_{\rm e}^2$. A down-scaling of $\dot M \propto 1/\sqrt D$ (where $D$
denotes the clumping factor) thus preserves the emission measure in the
atmosphere.  Because WR atmospheres are dominated by recombination processes,
the ionization structure as well as the emission line spectrum are,
approximately, preserved under such a transformation \citep{ham1:98}. The same
holds for the radiative {\em force}. The resulting radiative {\em
  acceleration} thus scales with $\sqrt D$ under this transformation.

\section{Spectral analyses of galactic WR stars}

\begin{figure}[t!]
\parbox[b]{0.99\textwidth}{\includegraphics[scale=0.65]{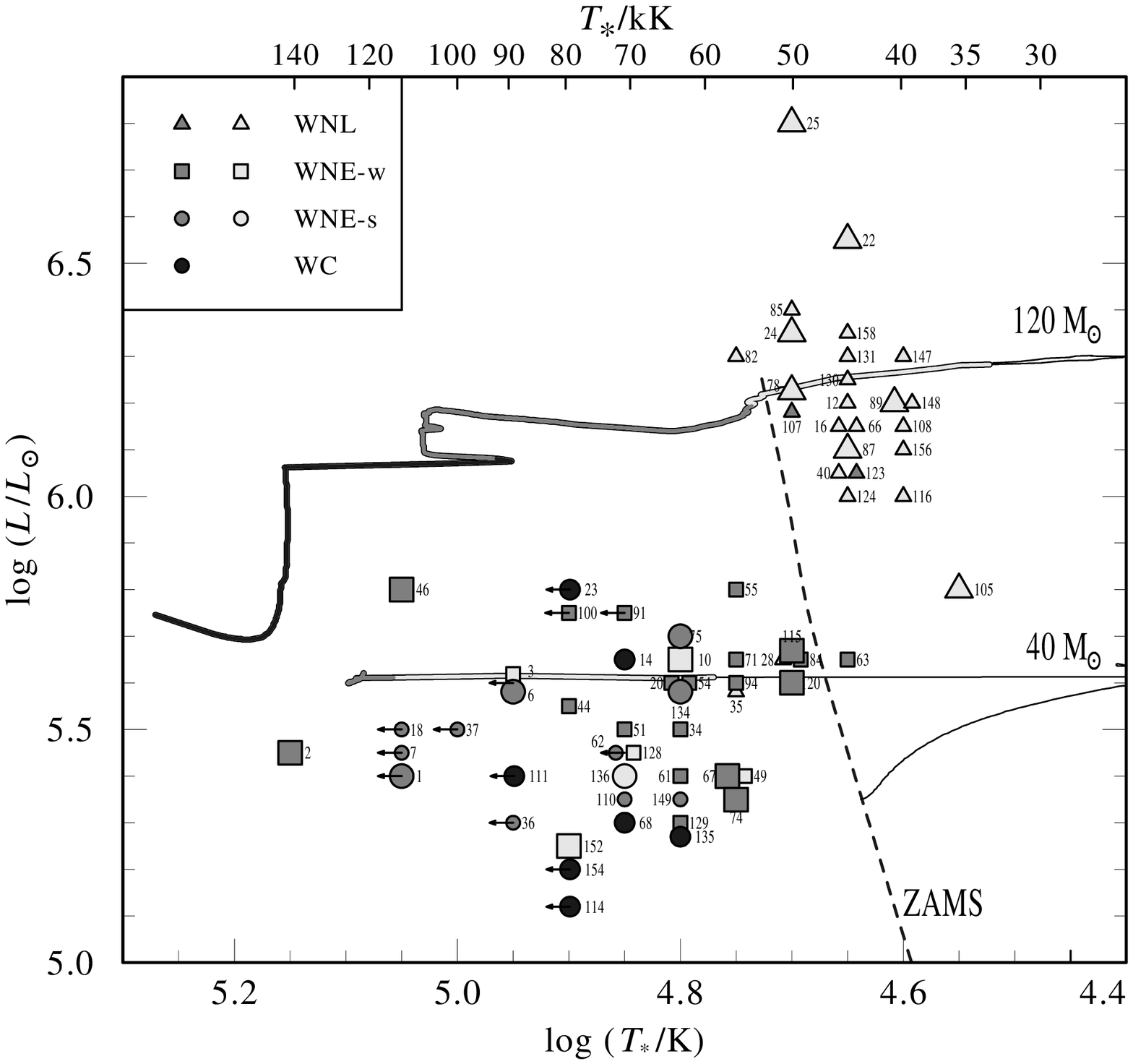}}
\caption{Recent spectral analyses of galactic WR stars with line-blanketed
  models, according to \citet{ham1:06} and \citet{bar1:06}: symbols in light
  grey denote H-rich WR stars, whereas H-free objects are indicated in dark
  grey.  WC stars are indicated in black.  For objects with large symbols
  direct distance estimates are available \citep{huc1:01}, whereas objects
  with small symbols are calibrated by their spectral subtype.  Evolutionary
  tracks for non-rotating massive stars \citep{mey1:03} are shown for
  comparison.
  \label{fig:hrd}
}
\end{figure}

Before we discuss the results from our hydrodynamic models we give an overview
of recent spectral analyses of galactic WR stars, based on line-blanketed PoWR
models. In Fig.\,\ref{fig:hrd} we show an empirical HRD with stellar
temperatures and luminosities taken from \citet[][WN stars]{ham1:06} and
\citet[][WC stars]{bar1:06}. According to these new analyses, the galactic WR
stars form two distinct groups in the HR diagram, which are distinguished by
their luminosities.  The first group is formed by H-rich WNL\,stars with
luminosities above $10^6\,L_\odot$. These stars are found to the right of the
ZAMS. The second group are early- to intermediate subtypes with lower
luminosities and hotter temperatures. The majority of these objects is found
to be H-free.

The observed dichotomy among the WN subtypes implies that the H-rich
WNL\,stars are the descendants of very massive stars, possibly still in the
pase of central H-burning.  The earlier subtypes (including the WC stars) are
less massive, He-burning objects. Note that distances are only known for a
small part of the WNL sample. Some of these objects might thus have lower
luminosities and be the direct progenitors of the earlier subtypes.

A fundamental problem arises from the frequent detection of H-free WR stars of
intermediate spectral subtypes, with stellar temperatures of $T_\star =
50$--$100\,$kK.  Due to their high mean molecular weight, these stars are
actually expected to lie on the He main-sequence with effective core
temperatures above $100\,$kK.  Although the new spectral analyses with
line-blanketed models tend to give higher values of $T_\star$, this
long-standing problem still persists.  Note that for a part of these stars the
wind is so thick that the effective photosphere lies far above the
hydrostatic radius.  For these cases, $T_\star$ which is the effective
temperature related to $R(\tau_{\rm Ross}=20)$ only gives a lower limit for
the hydrostatic core temperature.

\section{Hydrodynamic models for early spectral subtypes}

\begin{figure}[t!]
\parbox[b]{0.5\textwidth}{\includegraphics[scale=0.482]{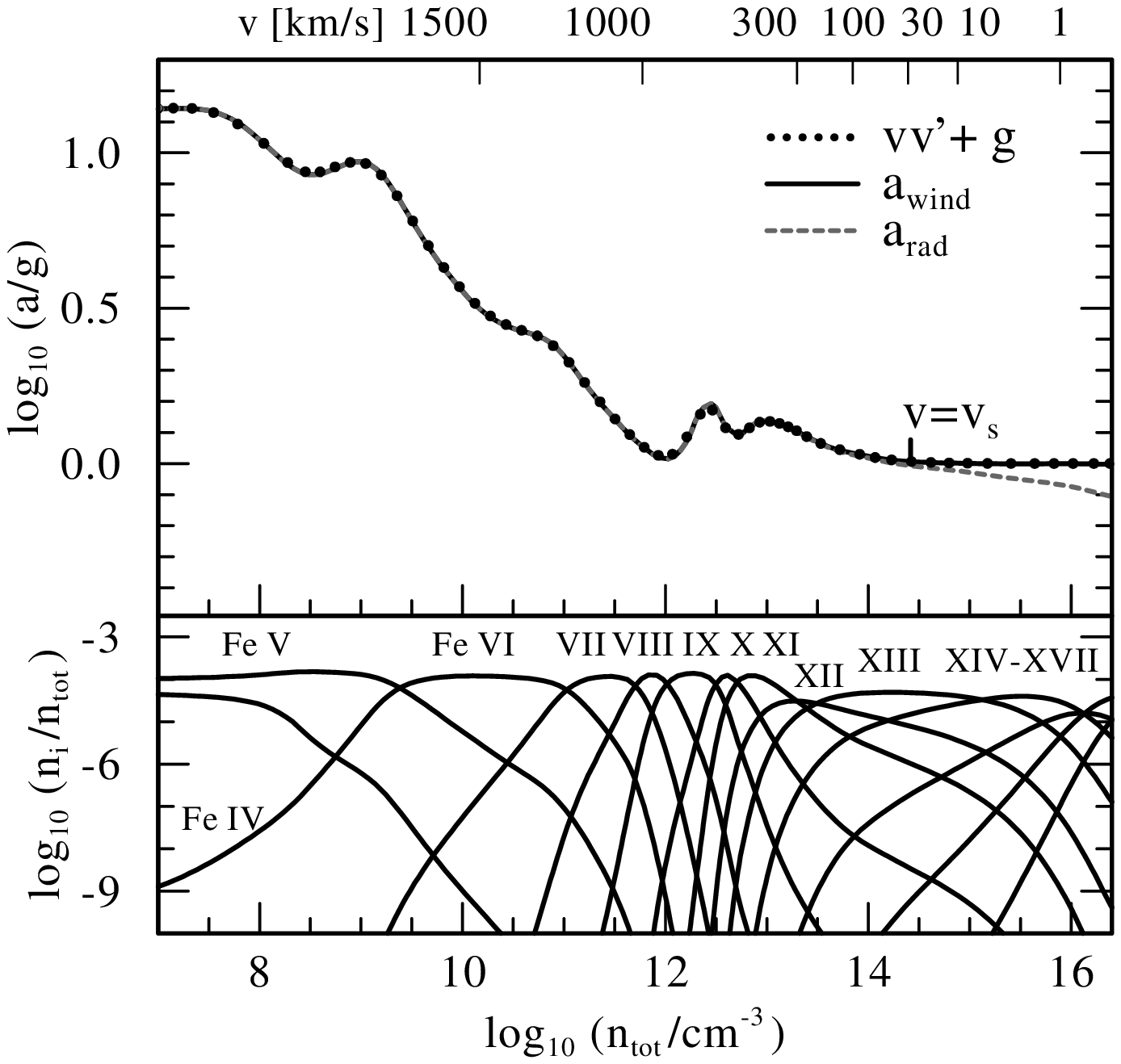}}
\parbox[b]{0.5\textwidth}{\includegraphics[scale=0.482]{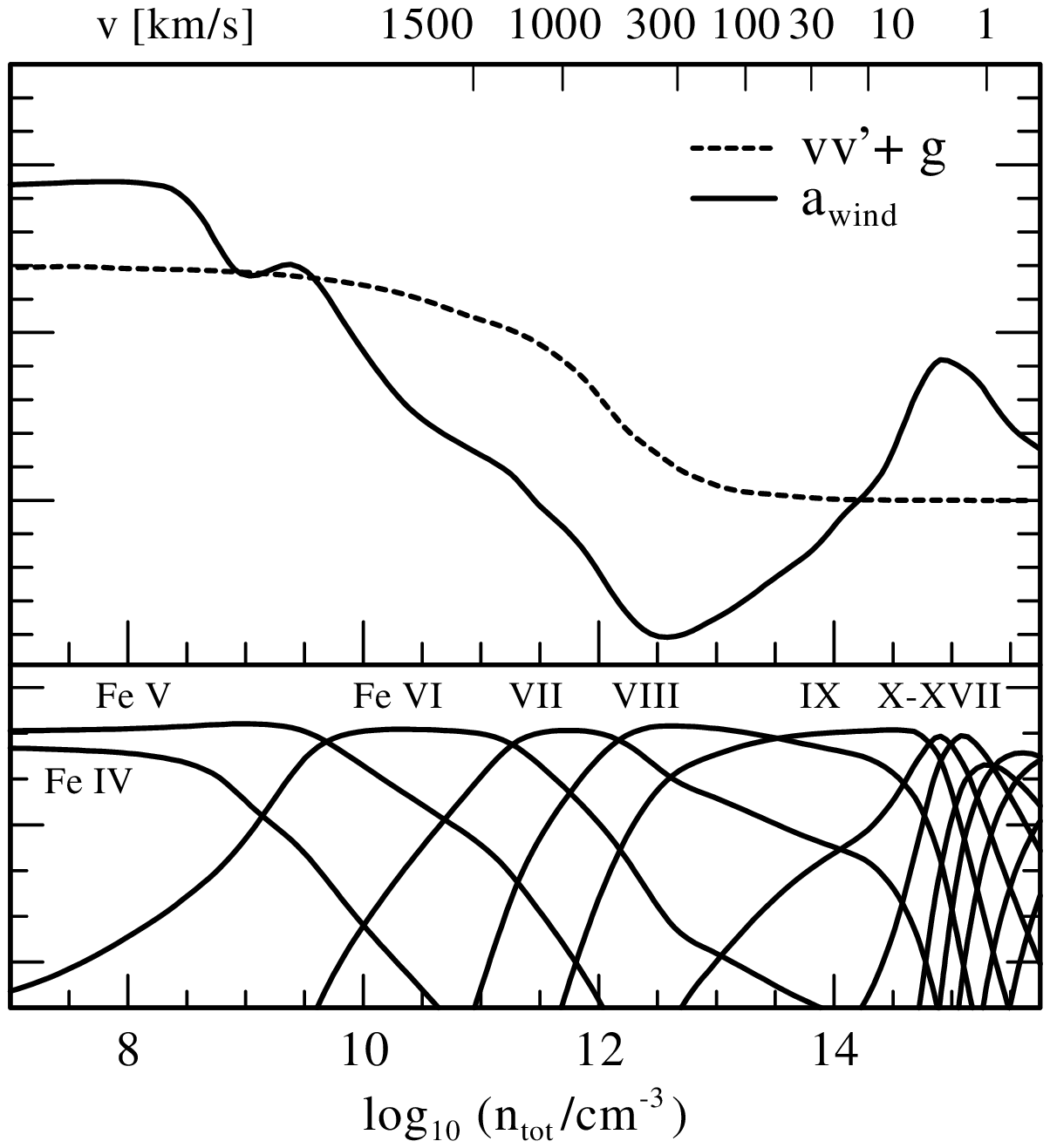}}
\caption{Wind acceleration within our models in units of the local gravity,
  vs.\ wind density as depth index (compare Fig.\,\ref{fig:3d}).  The left
  panel shows our hydrodynamic WC model with $T_\star=140\,$kK.
  Self-consistency is obtained throughout the complete atmosphere ($a_{\rm
    wind}=vv'+g$). The right panel shows a cooler model with $T_\star=85\,$kK.
  For this model no consistency is obtained. The hot Fe-peak opacities
  (Fe\,{\sc ix}--{\sc xvii}) are located so deep inside the atmosphere, that
  they are missing in the wind acceleration region.
   \label{fig:acc}
}
\end{figure}

As mentioned earlier, our hydrodynamic model atmospheres are capable to
reproduce the observed properties of early-type WC stars in a self-consistent
manner \citep{gra1:05}. In agreement with \citet{nug1:02} we find that high
critical-point temperatures ($\approx 200\,$kK) are needed to match the region
where the ``hot Fe opacity peak'' becomes active. According to our models,
this also requires extremely high stellar temperatures of $T_\star > 100$\,kK.
Otherwise, as illustrated in Fig.\,\ref{fig:acc}, the hot Fe-peak moves so
deep inside the atmosphere that no line opacities are available for the
acceleration of the wind base.  For intermediate spectral subtypes with
$T_\star = 70\,$--100\,kK our models thus predict no WR-type mass loss.  The
existence of H-free WR stars in this temperature range is thus ruled out by
the theory of stellar structure {\em and} our wind models.

Nevertheless, we have seen that such objects are commonly observed (see
Fig.\,\ref{fig:hrd}). The additional extension of their envelopes, as well as
their strong mass loss, can not be explained by standard assumptions. We thus
conclude that other processes might be important for these objects. In
particular, this means that their stellar winds might not be purely
radiatively driven. According to our models the lack of wind acceleration
occurs at the bottom of the stellar wind. Fast rotation seems to be a good
candidate for an additional support of the wind in this region.

\section{Hydrodynamic models for late spectral subtypes}

\begin{figure}[t!]
\begin{center}
\parbox[b]{0.75\textwidth}{\includegraphics[scale=0.45]{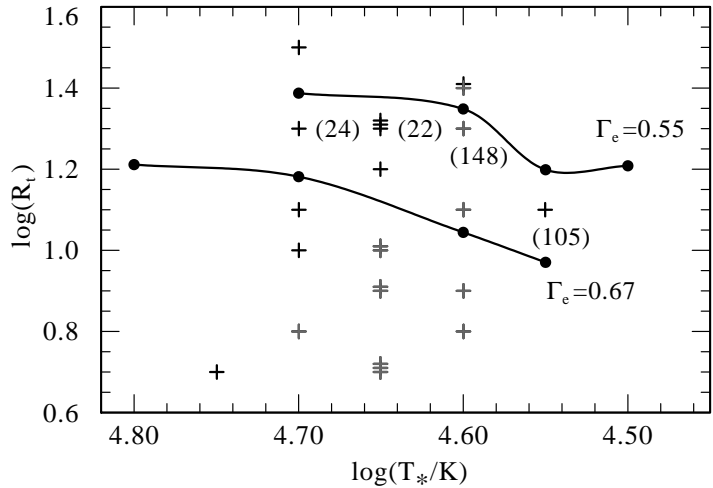}}
\end{center}
\caption{Wind models for galactic WNL stars: wind densities compared
  to observed values according to \citet{ham1:06}.  In the $R_{\rm
    t}$--$T_\star$ plane, our models (solid curves) reproduce the {\em upper}
  part of the observed WNL sample (crosses), corresponding to WNL stars with
  {\em low} wind densities (see text for the definition of $R_{\rm t}$). The
  objects belonging to the enigmatic class of WN\,8 spectral subtypes are
  indicated in grey.
  \label{fig:rtts}
}
\end{figure}

In a recent work we have investigated the properties of luminous, H-rich WNL
stars with hydrodynamic PoWR models \citep{gra2:06}. The most important
conclusion from that work is that WR-type mass loss is primarily triggered by
high $L/M$ ratios or, equivalently, Eddington factors $\Gamma_{\rm
  e}\equiv\chi_{\rm e}L_\star/4\pi c G M_\star$ approaching unity. Due to the
increase of the density scale height at low effective gravities, the formation
of optically thick winds seems to be strongly favored for such objects.  Note
that high $L/M$ ratios are expected for very massive stars {\em and} for
He-burning objects, giving a natural explanation for the occurrence of the
WR-phenomenon. 

In Fig.\,\ref{fig:rtts} we show a quantitative comparison of the results from
our hydrodynamic models with observations of the galactic WNL sample
\citep[according to][]{ham1:06}.  For this comparison we employ the
transformed radius
\begin{equation}
\label{eq:rtrans}
R_{\rm t} = R_\star \left[\frac{{v}_\infty}{2500 \, {\rm km}\,{\rm s^{-1}}} 
\left/
\frac{\sqrt{D}\dot{M}}{10^{-4} \, {\rm M_\odot}\,{\rm yr^{-1}}}\right]^{2/3} 
\right.  ,
\end{equation}
which is a luminosity-independent measure of the wind density. Models with the
same $R_{\rm t}$ are scale invariant, and show a very similar spectral
appearance for given $T_\star$ and given surface chemistry
\citep[see][]{ham1:98}. In the $R_{\rm t}$-$T_\star$ plane, our models with
Eddington factors $\Gamma_{\rm e}=0.55$ and 0.67 reproduce the observed WNL
stars with large $R_{\rm t}$. These are the WNL stars with relatively {\em
  low} wind densities. The resultant mass loss rates depend strongly on the
stellar temperature (with $\dot M \propto T_\star^{-3.5}$), nicely reflecting
the observed sequence of spectral subtypes from WN\,6 to WN\,9.

The high wind density objects, however, are not reproduced by our models.
Notably, these objects are dominated by the enigmatic WN\,8 subtypes, which
tend to show strong emission lines in combination with photometric variations
\citep[e.g.][]{mar1:98,lef1:05}.  Because of the apparent dependence of $\dot
M$ on $\Gamma_{\rm e}$ in Fig.\,\ref{fig:rtts}, one might think that higher
mass loss rates can easily be obtained by a further increase of $\Gamma_{\rm
  e}$. In our models, however, this leads to a point where the radiative
acceleration exceeds gravity in the hydrostatic layers.

For the relatively cool WNL stars this situation occurs due to the hot Fe-peak
which causes an inward increase of the Rosseland mean opacity from a certain
point on.  If gravity is exceeded by the resulting radiative force, a
stationary solution can only be obtained if the excess acceleration is
compensated by an outward increase of the gas pressure, i.e.\ a density
inversion.  In reality this will presumably lead to a non-stationary
situation.  \citet{dor1:06} have recently shown that the observed variability
in WN\,8 subtypes can indeed be explained by instabilities which are caused by
the hot Fe-peak. The obtained timescales around 0.5\,d are in rough agreement
with the observed period of 9.8\,h for WR\,123 \citep{lef1:05}, and with the
dynamical timescale of the wind ($R_\star/v_\infty \approx 4\,$h).  We thus
conclude that the observed pulsations will most probably influence the mass
loss of these objects, i.e., their winds are not purely radiation driven.
Note that the non-detection of X-ray emission from WN\,8 subtypes
\citep{osk1:05} might also be a hint on a different wind driving mechanism for
these objects.

Another reason for our difficulties with the WN\,8 subtypes might be related
to the fact that the distances of these objects are only poorly known.  They
might thus fall into the category of core He-burning post-RSG stars with much
lower luminosities. A detailed investigation of the influence of $L_\star$ on
our wind models, however, remains a topic of future investigations.

Finally, due to their proximity to the Eddington limit, our models allow to
estimate the masses of WNL stars. Because of the strong dependence of $\dot M$
on the $L/M$ ratio, we can give reliable mass estimates, that only depend on
the adopted distance to the object. For the H-rich, weak-lined WNL stars our
models tend to give very high masses, in agreement with very massive stars in
a late stage of the H-burning phase. For the case of WR\,22, a weak-lined
WN\,7h subtype with $L_\star = 10^{6.3}L_\odot$, we derive a stellar mass of
$78\,M_\odot$.  This value is in agreement with the work of \citet{rau1:96},
who derived $M_\star\,\sin^3 i = (72\pm 3)\,M_\odot$ from orbital elements.

\section{Conclusions}

Our hydrodynamic atmosphere calculations for WR\,stars have shown that WR-type
stellar winds can be driven alone by radiative acceleration.  In particular we
found that the proximity to the Eddington limit is the primary trigger of the
enhanced mass loss of WR\,stars. Nevertheless, difficulties remain with H-free
objects of intermediate spectral subtype, and late-type WN stars with strong
stellar winds (mostly of spectral type WN\,8). In both cases, problems are
caused by the gap between the 'hot' and the 'cool' Fe opacity-peaks. For the
intermediate subtypes this gap causes a deficiency of line opacities at the
wind base. This problem could be overcome by an additional support due to fast
stellar rotation. For the WN\,8 stars, the inward increase of the radiative
force due to the hot Fe opacity-peak causes problems in the hydrostatic
layers, presumably leading to the observed variability. These oscillations
most probably affect the mass loss of these stars.

\newpage


\end{document}